\documentclass[twocolumn]{emulateapj}

\usepackage{times}
\usepackage{amsmath,amssymb,amstext}
\usepackage[colorlinks=blue,linkcolor=blue,urlcolor=blue,citecolor=blue]{hyperref} 
\usepackage[all]{hypcap}
\usepackage{tabularx}
\usepackage{float}
\usepackage{natbib}
\usepackage{longtable}
\usepackage[usenames, dvipsnames]{color}
\usepackage[switch]{lineno}

\begin{document}

\title{Nustar view of the central region of the Perseus cluster}
\author{B. Rani\altaffilmark{1}, 
  G. M. Madejski\altaffilmark{2}, 
  R. F. Mushotzky\altaffilmark{3},
  C. Reynolds\altaffilmark{4},
  J. A. Hodgson\altaffilmark{5}
}
\affil{$^1$NASA Goddard Space Flight Center, Greenbelt, MD, 20771, USA}
\affil{$^2$ Kavli Institute for Particle Astrophysics and Cosmology, SLAC and Stanford University, Stanford, CA 94305 }
\affil{$^3$ Department of Astronomy, University of Maryland, College Park, MD 20742, USA}
\affil{$^4$ Institute of Astronomy, University of Cambridge, Madingley Road, Cambridge, CB3 OHA, UK}
\affil{$^5$ Korea Astronomy and Space Science Institute, 776 Daedeokdae-ro, Yuseong-gu, Daejeon, 30455, Korea}
\thanks{NPP Fellow}
\email{bindu.rani@nasa.gov}
\begin{abstract}

Located at the center of the Perseus cluster, 3C~84 is an extremely  bright and nearby radio galaxy. Because 
of the strong diffuse thermal emission from the cluster in X-rays, the detailed properties and the origin  
of a power-law component from 
the central active galactic nucleus (AGN)   
remains unclear in the source. We report here the first {\it NuSTAR} observations of 3C~84. The source was observed 
for 24.2 and 32~ks on February 01 and 04, 2018, respectively. {\it NuSTAR} observations spectrally decompose the 
power-law AGN component above 10~keV. The power-law component 
dominates the spectrum above  20~keV with a photon index $\sim$1.9 and an energy flux  F$_{20-30~keV}$ = 
1.0~$\times$10$^{-11}$~erg~cm$^{-2}$~s$^{-1}$, corresponding to an isotropic luminosity, 
$L_{20-30~keV}$ =  7.4$\times$10$^{42}$~erg~s$^{-1}$. We discuss possible emitting sites  
for the power-law component. The expected thermal 
emission from the accretion disk is not hot enough to account for the hard X-rays detected from the source.
Similar X-ray and $\gamma$-ray photon indices and long-term flux variations, the absence of 
cutoff energy in the hard X-ray spectrum of the source, correlated hard X-ray flux and hardness ratio 
variations, and the similarity of optical-X-ray slope to blazar rather than Seyfert galaxies 
supports the  hard X-ray power-law component originating from the jet.   

\end{abstract}


\keywords{galaxies: active -- quasars: individual: 3C~84 -- 
             X-rays  
               }


\section{Introduction}
The Perseus cluster of galaxies is one of the  X-ray brightest and nearest clusters in the
sky. At the center of the Perseus cluster, 3C~84 is a nearby (z=0.0172)  
radio loud active galactic nucleus (AGN) with a black hole mass $\sim$(3.4--8.0)$\times$10$^{8}$~M$_{\odot}$ 
\citep{wilman2005, scharwachter2013} and a relatively low Eddington ratio, 
$L_B/L_{Edd}$ $\sim$ 3$\times10^{-4}$ \citep{sikora2007}.  
The source is classified as a Seyfert Type 1.5 galaxy \citep{ho1997}.
Being at the center of a bright cluster, the X-ray properties of 3C~84 
have been extensively studied. \citet{fabian2015} compared the long-term X-ray variations 
with 90~GHz radio flux and found similar 
variations in the two.  
{\it XMM-Newton} observations taken in 2001 suggest the presence of a narrow Fe-K$\alpha$ 
line in the spectrum \citep{churazov2003}. The recent observations by the {\it Hitomi}/Soft X-ray 
telescope (SXS) 
confirm the detection of the 6.4~keV Fe-K$\alpha$ line at the 5.4$\sigma$ level, and suggest 
that the line 
originates within $\sim$1.6~kpc of the 3C~84 core \citep{hitomi_sxs}.

\begin{figure}
\centering
\includegraphics[scale=0.28,angle=0, trim=430 200 160 5, clip]{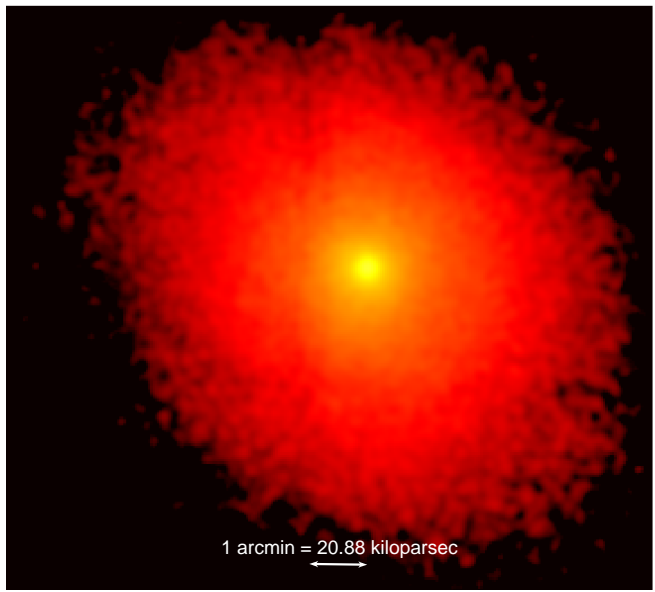}
\caption{{\it NuSTAR}  image of the Perseus 
cluster in the 3-70~KeV Kev band taken on Feb. 04, 2017. 
The bright 
emission at the center of the image is  3C 84.  }
\label{plot_fig1}
\end{figure}

The source has 
been detected at GeV energies since the beginning of {\it Fermi} Large Area Telescope (LAT) 
observations \citep{abdo2009}.  
Being relatively  inactive 
throughout the 1990s, there has been  an increase in $\gamma$-ray flux and cm-radio activity, 
which appears to have begun in 2005 \citep{abdo2009, nagai2010, suzuki2012}.   
From December 2016 to early January 2017, the source was detected at TeV energies
\citep{mukherjee2017, mirzoyan2017, tev_magic}.  
An  increase in the radio, soft X-ray and $\gamma$-ray flux was also detected during this 
flaring activity period \citep{hodgson2018, fukazawa2018}.

The X-ray emission of the source is strongly dominated by the thermal diffuse radiation from the cluster and  
a power-law  component  associated with the central AGN \citep{hitomi_sxs, ajello2009}. However the 
physical origin 
 this power-law component, e.g., whether it is more similar to the jet-like origin of the X-ray continuum 
 in blazars or the thermal Comptonization component in Seyfert galaxies, is not clear.
Similarities in the X-ray and 
radio/$\gamma$-ray flux variations \citep{fabian2015, fukazawa2018} clearly hint in 
favor of the presence of a jet-based emission component.  
We present here the first {\it NuSTAR} observations 
of  the Perseus cluster, which spectrally decompose the power-law emission component 
from the central AGN, and present the {\it Swift}-BAT data 
revealing variability of the hard X-ray flux.
This paper is structured as follows. Observations and data 
reduction are described in Sect.~2. Analysis and results are given in Sect.~3, and   
Sect.~4 and 5 presents 
the discussion and  conclusions, respectively.


\section{Observations and data reduction}
At the end of 2016, 3C~84 
went through an extreme flaring activity at GeV/TeV energies
\citep{mirzoyan2017, hodgson2018}.  To follow up the flaring activity at hard 
X-rays, we requested two {\it NuSTAR} \citep{Harrison2013} 
observations of the source. {\it NuSTAR} observations were taken in 
February 2017 (observations IDs : 90202046002, 90202046004).  The pointing  
on February 01, 2017, resulted in 24.2~ks of observing time, while on 
February 04, 2017,  32~ks were obtained. We also use the archival 105~months of 
{\it Swift}/BAT data, obtained from \citet{bat_105};  we 
detail the spectral fitting of those data in Section \ref{corona}.

Data were analyzed using the {\it NuSTAR} Data 
Analysis Software (NuSTARDAS) package v.1.3.1. After processing the 
raw data via {\tt nupipeline}, we used {\tt nuproducts} to extract 
higher level products, i.e.\ source spectra and  light curves.  
A hard X-ray image of the source can be seen in Fig. \ref{plot_fig1}. 
Since the Perseus cluster is a spatially extended source, we used  different approaches 
to extract the flux and spectrum of the central AGN 3C~84. 
Data for the central AGN component were extracted from a region 
of $30"$ radius, centered on the position of 3C~84 
(RA=3:19:48.38,Dec.=+41:30:42.10). Background was extracted from a 
$30"$ radius region roughly $180"$ SW of the source location. 
We also cross-check the results 
by selecting an annular background region  centered at the 
source position with an inner radius of $200"$ and an outer radius 
of $230"$.   
We binned the spectra   in order to have at least 
30 counts per rebinned channel.  For spectral fitting, we considered the channels 
corresponding nominally to the 3--70~keV energy range, where the source 
was robustly detected. In either case, we treat the thermal cluster emission as a background, 
and do not care about the residual thermal cluster flux in the source extraction region, 
only about the hard X-ray emission dominating at higher energies.  Our implicit 
assumption is that the residual cluster emission at the 3C84 extraction region and 
at the region used by us as background, have roughly the same temperature \citep{X-ray_isothermal}.

\begin{figure}
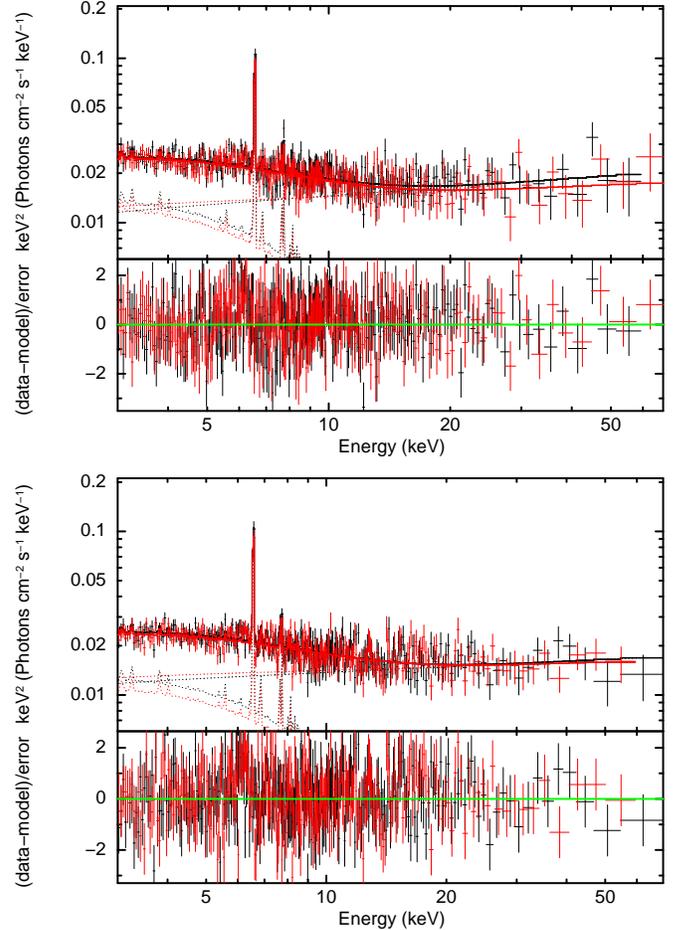

\centering
\includegraphics[scale=0.34,angle=-90, trim=0 30 0 0, clip]{grp_90202046002_final_apec_PL.ps}
\includegraphics[scale=0.34,angle=-90, trim=0 30 0 0, clip]{grp_90202046004_final_apec_PL.ps}
\caption{
Fitted {\it NuSTAR} spectrum of 3C~84 -- data taken on Feb. 01, 2017 (middle) and Feb. 04, 2017 (bottom). 
  Plus symbols are the observed data while the lines represent the fitted models. }
\label{plot_fig2}
\end{figure}

\begin{table*}
\label{tab1}
\caption{Parameters of the spectral fits }
\begin{tabular}{l l c c c c } \hline 
Model  & Parameter & \multicolumn{2}{c}{Period A}     & \multicolumn{2}{c}{Period B} \\
component       &    & FPMA         & FPMB            & FPMA        &  FPMB             \\\hline 
$wabs$ & nH (column density in units of $10^{22}$ cm$^{-2}$) & 0.14 (frozen) & & &  \\  
\smallskip 
$apec$& KT (plasma temperature in keV)      & 3.47$^{-0.42}_{+0.35}$           & --  & 3.70$^{-0.29}_{+0.34}$ & --   \\
\smallskip 
       &Abun (metal abundances)    & 0.39$^{-0.09}_{+0.19}$           & --  & 0.39$^{-0.11}_{+0.16}$ & --  \\
       \smallskip 
       & z (redshift)         &0.017 (frozen) & & & \\
       \smallskip 
       & Norm (apec normalization)    & 0.044$^{-0.011}_{+0.012}$        & 0.039$^{-0.011}_{+0.010}$    & 0.039$^{-0.007}_{+0.008}$  & 0.034$^{-0.008}_{+0.008}$  \\
       \smallskip 
$power-law$ & $\alpha$ (power-law index) & 1.85$^{-0.14}_{+0.12}$           & --  & 1.90$^{-0.13}_{+0.11}$ & -- \\
\smallskip 
& K (power-law normalization) &0.00974$^{-0.0033}_{+0.0049}$ &0.0114$^{-0.0032}_{+0.0048}$     & 0.0106$^{-0.0036}_{+0.0037}$  & 0.0120$^{-0.0035}_{+0.0037}$ \\
\smallskip 
$\chi^{2}$/dof (prob.) &reduced $\chi^{2}$/degrees of freedom (a corresponding probability) &0.89/454 (0.96)  & -- &1.03/564 (0.28)  &-- \\\hline 
\smallskip 
F$_{20-30~keV}$ & AGN flux in units of 10$^{-11}$ erg~cm$^{-2}$~s$^{-1}$ & 1.01$^{-0.05}_{+0.01}$  & 0.99$^{-0.03}_{+0.01}$ & 0.99$^{-0.37}_{+0.19}$ & 0.98$^{-0.20}_{+0.01}$ \\ 
\smallskip
F$_{2-10~keV}$ & AGN flux in units of 10$^{-11}$ erg~cm$^{-2}$~s$^{-1}$ & 3.36$^{-0.03}_{+0.03}$ & 3.5$^{-0.08}_{+0.08}$ & 3.20$^{-0.04}_{+0.05}$ & 3.40$^{-0.05}_{+0.06}$ \\\hline 
 & & & & & \\
   & COMBINED {\it NuSTAR} AND BAT SPECTRAL FITS& & & & \\\hline 
{\bf Model1:}    & Parameter    &      {\it NuSTAR} data        & BAT data   & &     \\\hline              
apec     &  kT (keV)    &        3.56$^{-0.25}_{+0.18}$             &  4.02$^{-0.31}_{+0.24}$   & &          \\
apec     &  norm        &        0.031$^{-0.002}_{+0.002}$      & 0.82$^{-0.13}_{+0.13}$    & &     \\
powerlaw &  PhoIndex    &        1.96$^{-0.04}_{+0.09}$          & --      & &        \\
powerlaw &  norm        &        0.013$^{-0.002}_{+0.002}$      & 0.008$^{-0.002}_{+0.003}$   & &   \\ 
$\chi^{2}$/dof (prob.) & &578/577 (0.20)   & -- \\\hline  
{\bf Model2:}     & Parameter    &      {\it NuSTAR} data        & BAT data   & &     \\\hline              
apec     &  kT (keV)    &        3.45$^{-0.33}_{+0.52}$             &  3.84$^{-0.73}_{+0.85}$        & &     \\
apec     &  norm        &        0.031$^{-0.003}_{+0.003}$     & 97$^{-70}_{+80}$     & &   \\
cutoffpl &  PhoIndex    &        1.94$^{-0.03}_{+0.03}$         & --         & &     \\
cutoffpl &  norm        &        0.013$^{-0.002}_{+0.002}$     & 0.008$^{-0.004}_{+0.003}$  & &    \\ 
cutoffpl & HighECut (keV) &     $>$100            &$>$100      & &       \\
$\chi^{2}$/dof (prob.) & &577/558 (0.21)   & -- & &\\\hline  
{\bf Model3:}   & Parameter    &      {\it NuSTAR} data        & BAT data    & &    \\\hline              
apec     &  kT (keV)    &        3.93$^{-0.33}_{+0.45}$            &  4.95$^{-0.94}_{+0.79}$     & &       \\
apec     &  norm        &        0.028$^{-0.002}_{+0.003}$     &  16$^{-17}_{+13}$   & &      \\
pexrav   &  PhoIndex    &        2.08$^{-0.21}_{+0.05}$          &  -- & &\\
pexrav   &  foldE (keV) &        $>$100     & $>$100 & &\\
pexrav   &  rel\_refl    &        0.50$^{-0.40}_{+0.45}$          & -- & & \\
pexrav   &  Redshift    &        0.017 (frozen)   & -- & & \\
pexrav   &  abund       &        1.0 (frozen)     & --  & & \\
pexrav   &  Fe\_abund    &        1.0  frozen)     & -- & & \\
pexrav   &  cosIncl     &        0.45 (frozen)    & -- & & \\
pexrav   &  norm        &        0.016$^{-0.003}_{+0.003}$      & 0.011$^{-0.004}_{+0.005}$ & & \\    
$\chi^{2}$/dof (prob.) & &575/558 (0.22)   & -- & & \\\hline  
\end{tabular}\\
The following parameters are Fixed  in {\bf Model1}, {\bf Model2}, and {\bf Model3}: 
nH/wabs (10$^{20}$ cm$^{-2}$)  = 0.14, Abundanc/apec = 0.50, Redshift = 0.017. 
We used the AtomDB (version 3.0.9) and the default abundance table is from \citet{anders1989}.\\
Note: The uncertainties on the parameters are 90$\%$ confidence limits.  \\
\end{table*}

\section{Analysis and Results} 
\label{result}

\subsection{Nustar data}
We used  $XSPEC$~V15.4.0 to model the source spectrum. 
In addition to an absorption model ($wabs$), which describes the absorption in the AGN by 
cold absorbing material in our Galaxy, we used a combined power-law, 
$A(E) = KE^{-\alpha}$,  and an emission spectrum from hot diffuse gas, $apec$,  to 
fit the source spectrum, i.e.\ $wabs*(apec+powerlaw)$. In $wabs$, we fixed the column density ($nH$)
to the Galactic value of 1.4$\times$10$^{21}$~cm$^{-2}$. The redshift ($z$) of the source 
is  fixed to 0.017. The plasma temperature ($kT$),
metal abundance ($abund$), and $apec$ normalization ($Norm$) are kept free while fitting the
spectrum.  Both the photon index ($\alpha$) and normalization ($K$) of the $power-law$ are
also set free. The model fit parameters and the reduced $\chi^{2}$ values along with
the degrees of freedom and the corresponding probability are given in Table \ref{tab1}. As
the statistics suggest, this simple $wabs*(apec+powerlaw)$ model well describes the source
spectrum over the energy range form  3.0 to 70.0~keV.

Motivated by the detection by the Hitomi satellite of the Fe K$\alpha$ line emission from 
the AGN \citep{hitomi_sxs}, we searched for a presence of such a line in the 
NuSTAR data.  The flux of the sum of the two components of the line measured by Hitomi 
is modest, at $\sim6 \times 10^{-6}$ photons cm$^{-2}$ s$^{-1}$.  We added a narrow Gaussian 
emission component to our combined $apec+powerlaw$ model, and note a significant improvement in 
$\chi^{2}$ (with $\Delta \chi^{2}$ of 10 for either observation).  We note that the best-fit 
of the line flux is significantly greater than that measured by Hitomi (by about a factor of 5, 
at $\sim30 \times 10^{-6}$ photons cm$^{-2}$ s$^{-1}$).  However, the fitted energy (as observed) 
is not well determined (values between 6.0 and 6.4 keV are allowed).  In conclusion, we consider as 
a possibility  to account for this finding the fact that the weak Fe-K line is sitting on the 
wing of the very strong (but  intrinsically narrow) 6.7 keV (rest) emission line from the hot 
cluster gas, and our detection is an artifact of slight miscalibration 
of NuSTAR's spectral response (specifically, the low-energy wing) to a narrow line.

A significant dominance of the thermal diffuse emission (see Fig.~\ref{plot_fig2}) in the source 
spectrum is seen up to $\sim$10~keV. 
The power-law AGN component clearly takes over above 10~keV. Most of the radiation
we see above 20~keV is contributed by the central AGN, 3C~84. In both
epochs of our observations, the
photon index is determined to be 1.8--1.9, which is quite similar to the 
$\gamma$-ray photon index during that period, $\Gamma_{E>100~MeV}$ $\sim$2.0 \citep{hodgson2018}. The 20-30~keV 
energy flux range is
0.9--1.1~$\times$10$^{-11}$~erg~cm$^{-2}$~s$^{-1}$, corresponding to  an
apparent isotropic X-ray luminosity, $L_{20-30~keV}$, of (7.35$\pm$0.66) $\times$10$^{42}$~erg~s$^{-1}$. 
We do not observe any variation in the
photon index or flux across the 4~day gap of  the two {\it NuSTAR} observations (see Table \ref{tab1}).

The estimated  soft X-ray, 2-10~keV, flux of the central region 
(including both the AGN and the residual cluster emission in the 30~arcsec radius 
extraction region) is  
$\sim$6$\times$10$^{-11}$~erg~cm$^{-2}$~s$^{-1}$. 
Soft X-ray flux of the AGN component  is $\sim$3.3$\times$10$^{-11}$~erg~cm$^{-2}$~s$^{-1}$ 
($L_{2-10~keV}$ = $\sim$2.3$\times$10$^{43}$~erg~s$^{-1}$), 
 which is similar to  the
AGN continuum flux, F$_{2-10~keV}$ $\sim$(1.96--4.36)$\times$10$^{-11}$~erg~cm$^{-2}$~s$^{-1}$,  
measured by the {\it Hitomi}/SXS observations in February, 2016 \citep{hitomi_sxs}. 
The  photon index is  also similar for the two observations. {\it Hitomi}/SXS measured the power-law index in the 
range 1.45-2.06, and for the {\it NuSTAR} observations, it is $\sim$1.9. 
The improved constraints relative to Hitomi result from NuSTAR's superior hard X-ray coverage 
(with sensitivity out to 70keV) and spatial resolution.

\subsection{Swift/BAT data} 
We used the archival {\it Swift}/BAT data  over  the 
first 105~months of observations from December 2004 to August 2013 \citep{bat_105} to investigate 
the hard X-ray spectral properties of the source.   
The detailed analysis of the BAT data can be found in \citet{bat_105}. 
We 
performed a combined fit of the {\it NuSTAR} and BAT data, which we discuss in Section \ref{corona}.
Note that the {\it NuSTAR} and BAT observations are not close in time and the BAT data are averaged 
over 9~years.

\section{Discussion}
\label{x-ray_location}

Given its high sensitivity and angular resolution at hard X-rays, {\it NuSTAR} observations spectrally decompose 
the power-law component in the 3C 84 spectrum. 
The power-law component of the spectrum could either be from the immediate vicinity, 
i.e.\ disk/corona, of the central black hole or from the jet. 
In the following sub-sections 
we discuss these possibilities in detail.

\subsection{Accretion disk}
For a 3-8$\times$10$^{8}$ M$_{\odot}$ black hole accreting at 10$\%$ of its $Eddington$ 
rate, the peak temperature at 1~$R_S$ is $\approx$2.5$\times$10$^{5}$~K, corresponding 
to 5.2$\times$10$^{15}$~Hz, which falls in the ultraviolet (UV) band.  Of course for 3C~84, the 
accretion  rate is way below 10$\%$ and $T$ $\propto$($\dot{L}$$_{Edd}$)$^{1/4}$; so, the disk would be 
even cooler. Moreover, the thermal radiation from the disk does not follow a power law.

\subsection{Corona}
\label{corona}
Inverse-Compton scattering of photons from the accretion disk by the corona could emit in hard 
X-rays \citep{reis2013, fabian2015}.  
For radiatively compact emission regions \citep[like corona, $\sim$2-30~$R_g$,][]{fabian2009, wilkins2011, sanfrutos2013},  
pair production naturally occurs. 
At extremely high temperature, pair   production can be a runway process, limiting any further rise
in the temperature and a producing a cutoff in the spectrum, which for Seyfert galaxies is in the 50-200~KeV 
range \citep{fabian2015}. 
However at low mass accretion rates ($L \leq 10^{-4}~L_{Edd}$) the 
compactness of the corona will be low and so very high temperatures can be achieved 
before the pair-production condition is met.

In the {\it NuSTAR} spectrum extending up to 70~keV, we do not see any hint of a cutoff in the spectrum. 
To further examine the possibility of a cutoff at higher energies, we performed 
a combined fit of the {\it NuSTAR} and {\it Swift}/BAT observations.  
We noticed a  clear disagreement between {\it NuSTAR} and BAT data 
between $\sim$(15-40)~keV, which could be anticipated as the data are not simultaneous. 
Moreover, the BAT spectrum is extracted from a larger region (19.5~arcmin) compared to NuSTAR and is 
dominated by cluster emission up to $\sim$50~KeV \citep{ajello2009}, which could be 
possibly responsible for the discrepancy between the two. Because of this reason, we ignore the 
BAT data below 50~KeV in the joint NuSTAR-BAT spectral fitting.

\begin{figure}
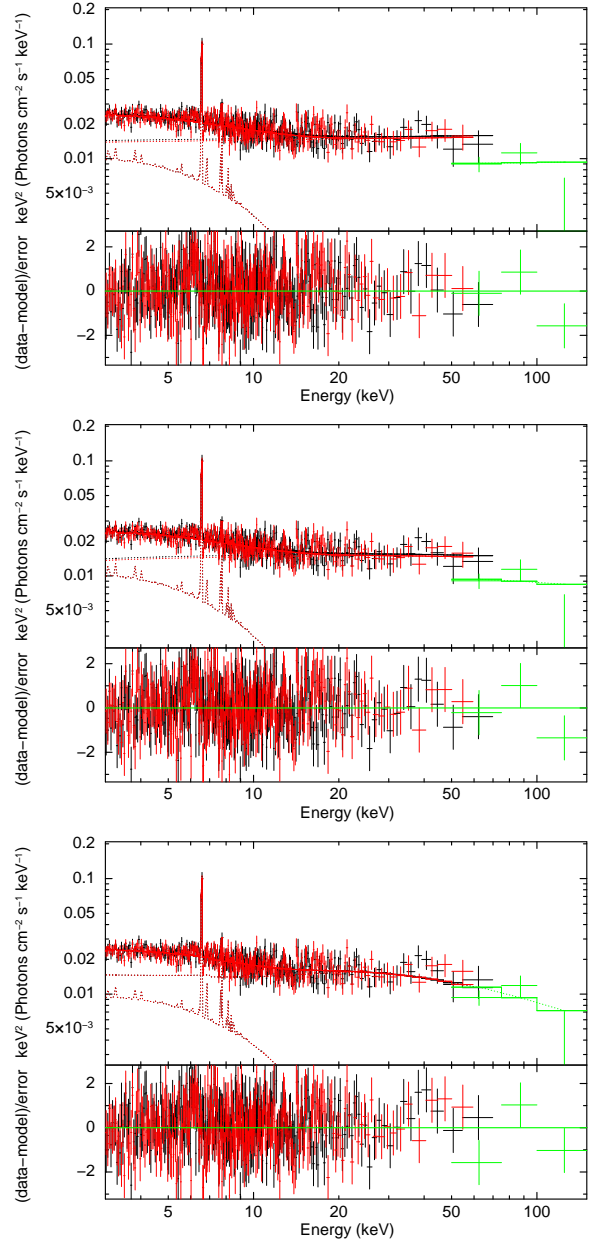

 \center
\includegraphics[scale=0.3,angle=-90, trim=0 0 0 0, clip]{model1.ps}
\\
\includegraphics[scale=0.3,angle=-90, trim=0 0 0 0, clip]{model2.ps}
\\
\includegraphics[scale=0.3,angle=-90, trim=0 0 0 0, clip]{model3.ps}
\caption{Combined {\it NuSTAR} and BAT spectrum fitted with {\bf Model1:} ($wabs*(apec+powerlaw)$, top), 
{\bf Model2:} ($wabs*(apec+cutoffpl)$, middle), and {\bf Model3:} ($wabs*(apec+pexrav)$, bottom). The 
best fit parameters are listed in Table \ref{tab1}.}
\label{nustar_bat_spec}
\end{figure}

We tried several models to fit the combined NuSTAR and BAT data. 
At first, we used the  simple $wabs*(apec+powerlaw)$ model ($Model1$) and froze the model parameters to their 
respective {\it NuSTAR} model-fit values except the $apec$ plasma temperature ($KT$) and normalization and the 
power-law index and its normalization. While we present here the spectral-fit results using the {\it NuSTAR} 
data taken on February 04, 2017, we obtained similar results for the first {\it NuSTAR} observation. 
The  model-fit parameters are given in Table \ref{tab1}, and the 
fitted spectrum is plotted in Fig.\ \ref{nustar_bat_spec} (top).   As the statistics suggest, $Model1$ well describes the source spectrum up to 200~KeV. Moreover, the 
power-law index for the combined {\it NuSTAR}-BAT data is  same as the power-law index for the {\it NuSTAR} 
spectral fits.

\begin{figure}
\includegraphics[scale=0.28,angle=0, trim=0 0 0 10, clip]{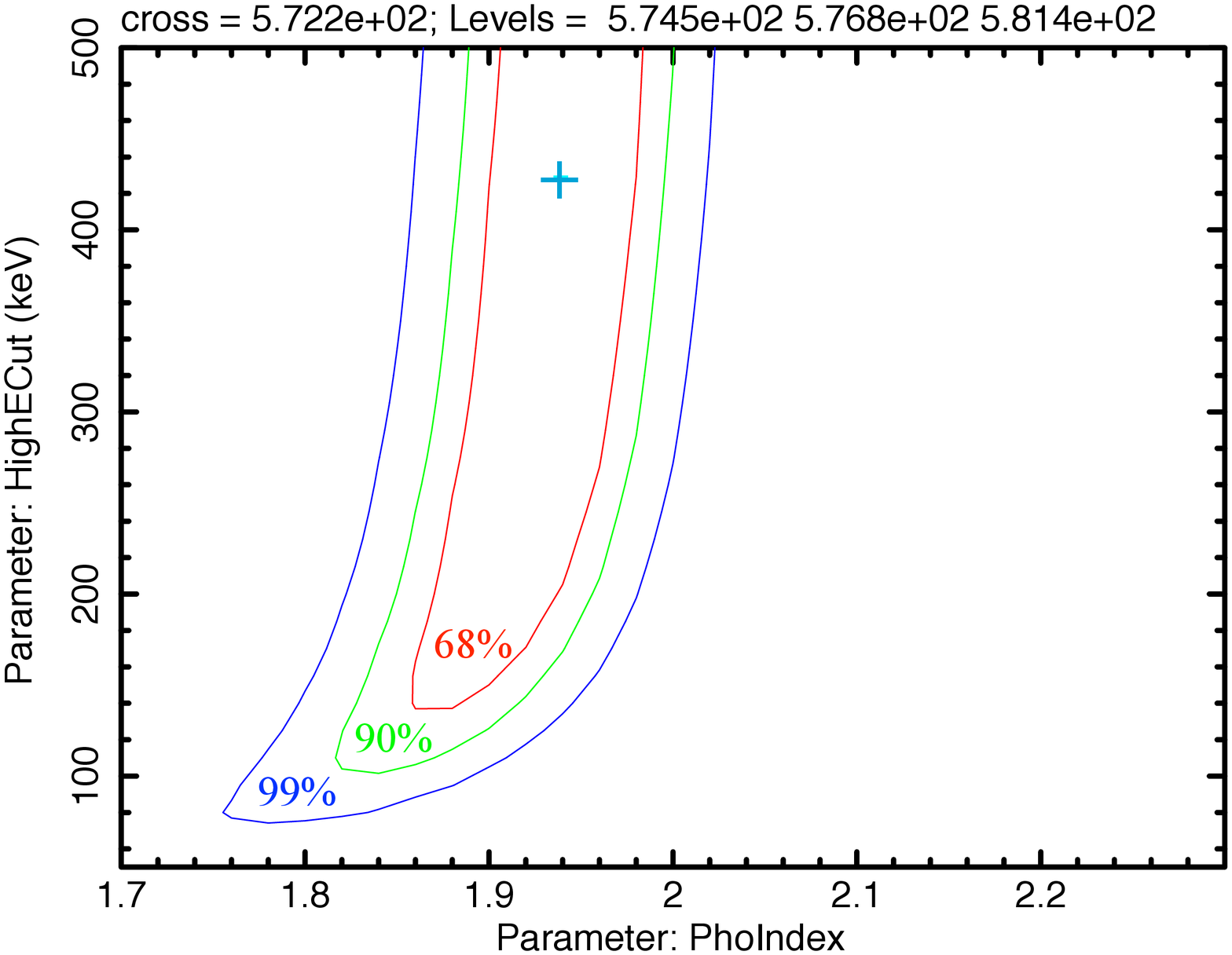}
\includegraphics[scale=0.28,angle=0, trim=0 0 0 10, clip]{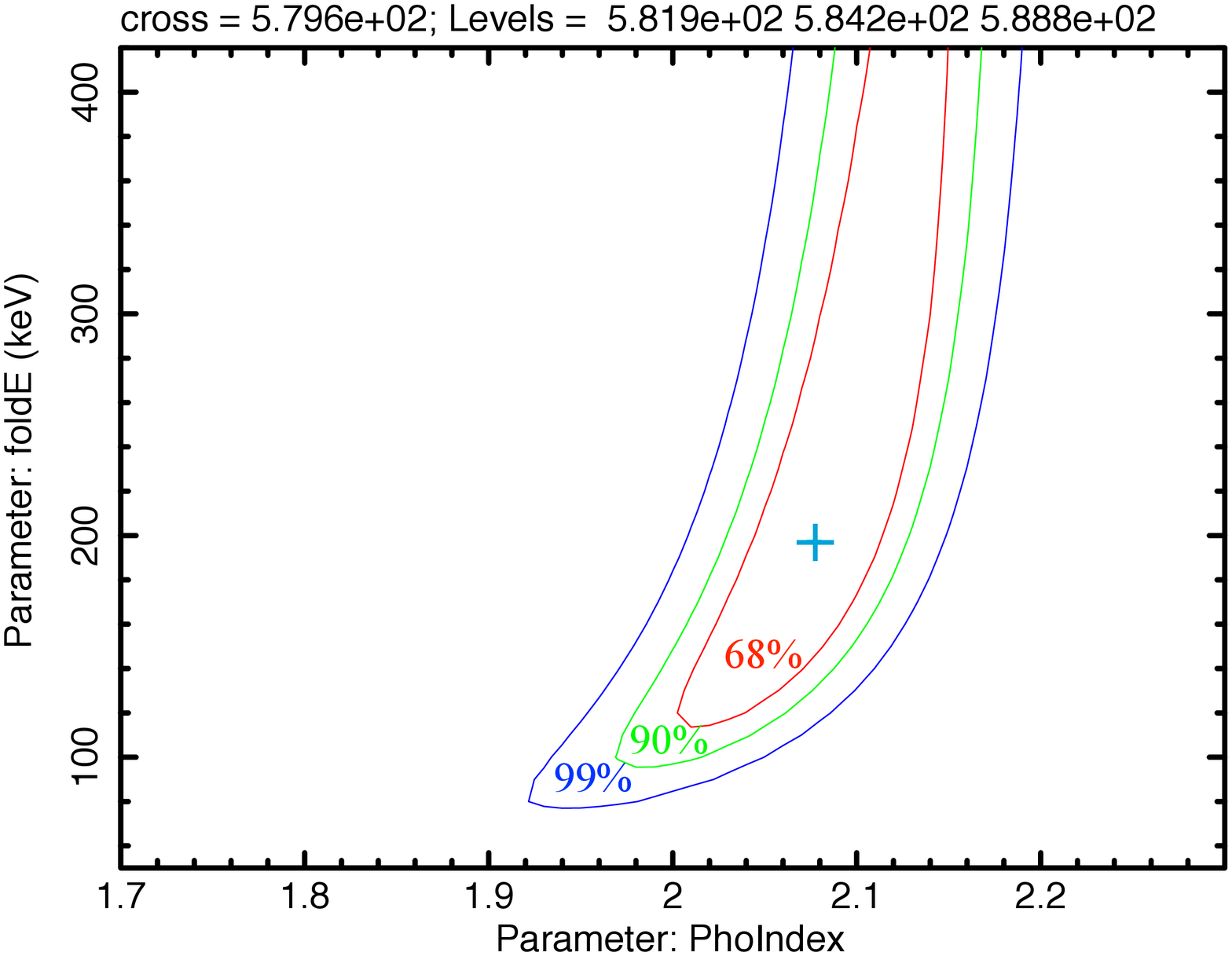}
\caption{{\it Top:} Confidence contours of cutoff energy versus photon index parameters in 
the $MODEL2$ (cutoff power-law, top) and $MODEL3$ (reflection, bottom). The best fit parameter 
(``+" symbol) and 68, 90, and 99$\%$ levels are shown.
}
\label{plot_cont}
\end{figure}

We next replaced the power-law component by a cutoff power-law 
($cutoffpl$) ($Model2$), and compared the test statistics of the two fits.  We noticed no 
significant change in the $\chi^{2}$ after adding the cutoff energy.   
Figure \ref{plot_cont} (top) illustrates the contour plots between the power-law photon index and 
cutoff energy, which  constrain the cutoff energy to be above 100~keV for values of the index between 
1.7-2.1. 
We further tested the reflection signatures in the spectrum \citep[$Model3$][]{reflection_mod}. 
Again no substantial improvement is noticed in the spectral fitting. 
The  best fit parameters are listed in Table \ref{tab1}, and the 
fitted spectrum is plotted in Fig.\ \ref{nustar_bat_spec}.
The index and cutoff energy contour plots in the reflection model are shown in 
Fig.\ \ref{plot_cont} (bottom). The reflection model constrained the cutoff energy 
to be above 100 keV for index 1.9-2.2. 
The absence of any significant improvement in the spectral fits by adding either cutoff energy or 
reflection component over a simple power-law component suggests that if there is a cutoff energy 
in the hard X-ray spectrum of the source, it is beyond the BAT energy range.  
Both spectral models constrain the cutoff energy to be above 100~keV, which argues in favor 
of the jet-based origin of the hard X-ray emission in the source.

\subsection{Jet}
\label{jet}
Since the photon index and flux behavior is quite similar in the {\it NuSTAR} X-ray and 
{\it Fermi}-LAT $\gamma$-ray observations, we used the {\it NuSTAR} spectral information to extrapolate and 
predict the $\gamma$-ray flux in the {\it Fermi} energy range using the 
$dummyrsp$\footnote{https://heasarc.gsfc.nasa.gov/xanadu/xspec/manual/node94.html} routine 
in $XSPEC$. The task returns a flux value of 1.9$\times$10$^{-10}$~erg~cm$^{-2}$~s$^{-1}$ 
corresponding to an isotropic luminosity  of $\sim$1.14$\times$10$^{44}$~erg~s$^{-1}$
in the 0.3--1.0~GeV energy range. The estimated isotropic $\gamma$-ray 
luminosity using {\it NuSTAR} observations is similar to the observed $\gamma$-ray luminosity 
of the source using the {\it Fermi}-LAT, i.e.\ L$_{\gamma-ray}$ $\sim$10$^{44}$~erg~s$^{-1}$ 
\citep{fukazawa2018, hodgson2018}. This suggests that the hard X-ray power-law 
spectrum  extends up to $\gamma$ rays. If this is true, 
the flux variations at the two energy bands should also have similar behavior. 
In fact, in a recent study \citet{fukazawa2018} found a similar variability trend in the 
long-term  5-10~keV X-ray and $\gamma$-ray flux variations. Moreover, the X-ray emission 
can be described as the low-energy tail of the inverse-Compton scattering in case 
of a one-zone synchrotron self-Compton model, suggesting a jet-based origin of the
X-ray variations \citep{fukazawa2018}. A significant correlation between $\gamma$-ray and 
radio flux variations suggests multiple high-energy emission sites located  within a 
distance of $\sim$2.0~parsecs from the central black hole \citep{hodgson2018}. 
Correlated X-ray and $\gamma$-ray variations in the source imply the same site for the 
X-ray emission, i.e.\ the power-law component of the X-ray emission is located within 
2~parsecs of the central 
black hole.

Another check for the jet-based origin of the hard X-ray
emission is the spectral slope from optical to X-rays ($\alpha_{OX}$). 
Seyfert galaxies (where most of the X-ray emission is 
from the immediate vicinity of the central black hole) have $<$$\alpha_{OX}$$>$ $\sim$1.4 
\citep{lusso2010}.  The slope could be a bit smaller for low-luminosity 
AGN like 3C~84 as $\alpha_{OX}$ is a function of luminosity \citep[see][for more details]{donato2001}.
For blazars (where X-ray emission is mostly from the jet),  $\alpha_{OX}$ is 
$\sim$1.0-1.2 \citep{donato2001}.
We compared the optical continuum flux at 1500$\AA$  from \citet{evans2004} to 
2-10~KeV X-ray flux, and the estimated   $\alpha_{OX}$ value for 3C~84 is 1.12, which is more 
similar to blazars than Seyfert galaxies favoring the jet-based origin of the X-ray emission.

Using the 
argument that X-ray and $\gamma$-ray emission in 3C~84 are co-spatial,  we  derive a lower 
limit on the  Doppler factor using \citep{dondi1995},
\begin{equation}
\delta \geq \Big [ 3.5 \times 10^3 \frac{(1+z)^{2\alpha} (1+z-\sqrt{1+z})^2 F_x (3.8 \nu \nu_x)^{\alpha}}{t_{var}} \Big ]^{1/(4+2\alpha)}
\end{equation}   
where $\nu_x$ is the X-ray frequency in keV and $F_x$ is the corresponding flux density in $\mu$Jy,   
$\alpha$ is the X-ray spectral index, $\nu$ is the $\gamma$-ray frequency, and $t_{var}$ is 
the variability timescale at high-energies (mostly used to constrain the size of the high-energy 
emission region). From the {\it NuSTAR} observations, we have $F_{20-30~keV}$ = 0.16~$\mu$Jy 
(flux density calculated from the X-ray flux measured in the 20-30~KeV band)
and $\Gamma_{20-30~keV}$ = 1.9. The source exhibits variations on multiple timescales 
at GeV energies, i.e.\ the rapid flares have doubling timescales of $\sim$10~hr 
\citep{tanada2018} while the long-duration outburst lasts for 8-12 months 
(250-350~days) \citep{hodgson2018}. 
Using $\nu_{X-ray}$= 25~keV, $\alpha_{X-ray}$ = 0.9 ($\Gamma_{20-30~keV}$ $-$ 1), and 
$t_{var}$ = 10~hr, we obtained $\delta_{1GeV}$ $\geq$ 1.4 and  $\delta_{1TeV}$ $\geq$ 3.9. 
Using $t_{var}$ = 300~days,  we have
$\delta_{1GeV}$ $\geq$ 0.4 and  $\delta_{1TeV}$ $\geq$ 1.3. 
The estimated Doppler factor values ($\delta \geq 0.4-3.9$) are  similar to those 
obtained using the broadband spectral energy 
distribution and jet kinematics \citep[$\delta \sim 1-3$,][]{fukazawa2018, hodgson2018}.

\section{Conclusion}
We present analysis of the {\it NuSTAR} observations of the central nucleus (3C~84) of the 
Perseus cluster. Because of the 
strong dominance of the diffuse thermal emission from the cluster, the true nature and origin of the power-law 
component from the central AGN remains under debate. The greater 
sensitivity at hard X-rays offered by {\it NuSTAR} allowed us to probe the AGN component in the 
source spectrum. A strong dominance of the thermal diffuse emission from the cluster is seen 
below 10~keV, while the power-law component takes over above 20~keV. 
No flux and spectral variations were observed in the two {\it NuSTAR} pointings  during 4 days. 
The 20-30~keV energy flux range is $\sim$~1.0$\times$10$^{-11}$~erg~cm$^{-2}$~s$^{-1}$ 
($L_{20-30~keV}$ = 7.4$\times$10$^{42}$~erg~s$^{-1}$) with a photon 
index of $\sim$1.9.

We discussed the possibility of different AGN components contributing to the power-law 
spectrum.  An accretion disk of a (3-8)$\times$10$^{8}$ M$_{\odot}$ black hole does not shine in X-rays. 
The following arguments suggest that the hard X-ray and $\gamma$-ray radiation is emitted by the same population 
of relativistic electrons from the jet: (1) X-ray and $\gamma$-ray 
photon indices are quite similar ($\Gamma$ $\sim$2), (2) $\gamma$-ray flux is just an extension 
of the X-ray flux, (3) the source exhibits similar long-term flux variations at X-rays and 
$\gamma$-rays, (4) no indication of cutoff energy is seen in the combined {\it NuSTAR} and BAT 
spectrum,   and 
(5)  the optical-to-Xray spectral slope is more similar to that of blazars rather than Seyfert galaxies. 
Our study therefore supports that the hard X-ray power-law component dominating the source spectrum above 20~KeV 
originates from the jet.

\acknowledgements
This research was supported by an appointment to the NASA Postdoctoral Program
at the Goddard Space Flight Center, administered by Universities Space Research Association 
through a contract with NASA. 
BR thanks Erin Kara, Mihoko Yukita, Dave Thompson, Stefan Walker, Amy Lien, Hans Krimm, 
and Chris Shrader,  for interesting discussions about  AGN disk and corona.


\begin{thebibliography}{}
\expandafter\ifx\csname natexlab\endcsname\relax\def\natexlab#1{#1}\fi

\bibitem[{{Abdo} {et~al.}(2009){Abdo}, {Ackermann}, {Ajello}, {Atwood},
  {Axelsson}, {Baldini}, {Ballet}, {Band}, {Barbiellini}, {Bastieri}, \&
  et~al.}]{abdo2009}
{Abdo}, A.~A., {Ackermann}, M., {Ajello}, M., {et~al.} 2009, \apjs, 183, 46

\bibitem[{{Ajello} {et~al.}(2009){Ajello}, {Rebusco}, {Cappelluti}, {Reimer},
  {B{\"o}hringer}, {Greiner}, {Gehrels}, {Tueller}, \& {Moretti}}]{ajello2009}
{Ajello}, M., {Rebusco}, P., {Cappelluti}, N., {et~al.} 2009, \apj, 690, 367

\bibitem[Anders \& Grevesse(1989)]{anders1989} Anders, E., \& Grevesse, N.\ 1989, \gca, 53, 197 


\bibitem[{{Churazov} {et~al.}(2003){Churazov}, {Forman}, {Jones}, \&
  {B{\"o}hringer}}]{churazov2003}
{Churazov}, E., {Forman}, W., {Jones}, C., \& {B{\"o}hringer}, H. 2003, \apj,
  590, 225

\bibitem[{{Donato} {et~al.}(2001){Donato}, {Ghisellini}, {Tagliaferri}, \&
  {Fossati}}]{donato2001}
{Donato}, D., {Ghisellini}, G., {Tagliaferri}, G., \& {Fossati}, G. 2001, \aap,
  375, 739

\bibitem[{{Dondi} \& {Ghisellini}(1995)}]{dondi1995}
{Dondi}, L., \& {Ghisellini}, G. 1995, \mnras, 273, 583

\bibitem[{{Evans} \& {Koratkar}(2004)}]{evans2004}
{Evans}, I.~N., \& {Koratkar}, A.~P. 2004, \apjs, 150, 73

\bibitem[{{Fabian} {et~al.}(2015){Fabian}, {Walker}, {Pinto}, {Russell}, \&
  {Edge}}]{fabian2015}
{Fabian}, A.~C., {Walker}, S.~A., {Pinto}, C., {Russell}, H.~R., \& {Edge},
  A.~C. 2015, \mnras, 451, 3061

\bibitem[{{Fabian} {et~al.}(2009){Fabian}, {Zoghbi}, {Ross}, {Uttley}, {Gallo},
  {Brandt}, {Blustin}, {Boller}, {Caballero-Garcia}, {Larsson}, {Miller},
  {Miniutti}, {Ponti}, {Reis}, {Reynolds}, {Tanaka}, \& {Young}}]{fabian2009}
{Fabian}, A.~C., {Zoghbi}, A., {Ross}, R.~R., {et~al.} 2009, \nat, 459, 540

\bibitem[{{Fukazawa} {et~al.}(2018){Fukazawa}, {Shiki}, {Tanaka}, {Itoh}}]{fukazawa2018}
{Fukazawa}, Y., {Shiki}, K., {Tanaka}, Y., {et~al.} 2018, \apj, 855, 93

\bibitem[{{Harrison} {et~al.}(2013){Harrison}, {Craig}, {Christensen},
  {Hailey}, {Zhang}, {Boggs}, {Stern}, {Cook}, {Forster}, {Giommi},
  {Grefenstette}, {Kim}, {Kitaguchi}, {Koglin}, {Madsen}, {Mao}, {Miyasaka},
  {Mori}, {Perri}, {Pivovaroff}, {Puccetti}, {Rana}, {Westergaard}, {Willis},
  {Zoglauer}, {An}, {Bachetti}, {Barri{\`e}re}, {Bellm}, {Bhalerao},
  {Brejnholt}, {Fuerst}, {Liebe}, {Markwardt}, {Nynka}, {Vogel}, {Walton},
  {Wik}, {Alexander}, {Cominsky}, {Hornschemeier}, {Hornstrup}, {Kaspi},
  {Madejski}, {Matt}, {Molendi}, {Smith}, {Tomsick}, {Ajello}, {Ballantyne},
  {Balokovi{\'c}}, {Barret}, {Bauer}, {Blandford}, {Brandt}, {Brenneman},
  {Chiang}, {Chakrabarty}, {Chenevez}, {Comastri}, {Dufour}, {Elvis}, {Fabian},
  {Farrah}, {Fryer}, {Gotthelf}, {Grindlay}, {Helfand}, {Krivonos}, {Meier},
  {Miller}, {Natalucci}, {Ogle}, {Ofek}, {Ptak}, {Reynolds}, {Rigby},
  {Tagliaferri}, {Thorsett}, {Treister}, \& {Urry}}]{Harrison2013}
{Harrison}, F.~A., {Craig}, W.~W., {Christensen}, F.~E., {et~al.} 2013, \apj,
  770, 103


  
\bibitem[Hitomi Collaboration et al.(2018)]{hitomi_sxs} Hitomi Collaboration, Aharonian, F., Akamatsu, H., et al.\ 2018, PASJ, 70, 13   

\bibitem[{{Ho} {et~al.}(1997){Ho}, {Filippenko}, \& {Sargent}}]{ho1997}
{Ho}, L.~C., {Filippenko}, A.~V., \& {Sargent}, W.~L.~W. 1997, \apjs, 112, 315

\bibitem[{{Hodgson} {et~al.}(2018){Hodgson}, {Rani}, {Lee}, {Algaba}, {Kino},
  {Trippe}, {Park}, {Zhao}, {Byun}, {Kang}, {Kim}, {Kim}, {Kim}, {Miyazaki},
  {Wajima}, {Oh}, {Kim}, \& {Gurwell}}]{hodgson2018}
{Hodgson}, J.~A., {Rani}, B., {Lee}, S.-S., {et~al.} 2018, \mnras, 475, 368

\bibitem[{{Lusso} {et~al.}(2010){Lusso}, {Comastri}, {Vignali}, {Zamorani},
  {Brusa}, {Gilli}, {Iwasawa}, {Salvato}, {Civano}, {Elvis}, {Merloni},
  {Bongiorno}, {Trump}, {Koekemoer}, {Schinnerer}, {Le Floc'h}, {Cappelluti},
  {Jahnke}, {Sargent}, {Silverman}, {Mainieri}, {Fiore}, {Bolzonella}, {Le
  F{\`e}vre}, {Garilli}, {Iovino}, {Kneib}, {Lamareille}, {Lilly}, {Mignoli},
  {Scodeggio}, \& {Vergani}}]{lusso2010}
{Lusso}, E., {Comastri}, A., {Vignali}, C., {et~al.} 2010, \aap, 512, A34

\bibitem[{{Magdziarz} \& {Zdziarski}(1995)}]{reflection_mod}
{Magdziarz}, P., \& {Zdziarski}, A.~A. 1995, \mnras, 273, 837

\bibitem[{{MAGIC collaboration} {et~al.}(2018){MAGIC collaboration}, {Ansoldi},
  {Antonelli}, {Arcaro}, {Baack}, {Babi{\'c}}, {Banerjee}, {Bangale}, {Barres
  de Almeida}, {Barrio}, {Becerra Gonz{\'a}lez}, {Bednarek}, {Bernardini},
  {Berse}, {Berti}, {Bhattacharyya}, {Bigongiari}, {Biland}, {Blanch},
  {Bonnoli}, {Carosi}, {Ceribella}, {Chatterjee}, {Colak}, {Colin}, {Colombo},
  {Contreras}, {Cortina}, {Covino}, {Cumani}, {D'Elia}, {Da Vela}, {Dazzi}, {De
  Angelis}, {De Lotto}, {Delfino}, {Delgado}, {Di Pierro}, {Dom{\'{\i}}nguez},
  {Dominis Prester}, {Dorner}, {Doro}, {Einecke}, {Elsaesser}, {Fallah
  Ramazani}, {Fattorini}, {Fern{\'a}ndez-Barral}, {Ferrara}, {Fidalgo},
  {Foffano}, {Fonseca}, {Font}, {Fruck}, {Galindo}, {Gallozzi}, {Garc{\'{\i}}a
  L{\'o}pez}, {Garczarczyk}, {Gaug}, {Giammaria}, {Godinovi{\'c}}, {Gora},
  {Guberman}, {Hadasch}, {Hahn}, {Hassan}, {Hayashida}, {Herrera}, {Hoang},
  {Hose}, {Hrupec}, {Ishio}, {Konno}, {Kubo}, {Kushida}, {Lamastra}, {Lelas},
  {Leone}, {Lindfors}, {Lombardi}, {Longo}, {L{\'o}pez}, {Maggio}, {Majumdar},
  {Makariev}, {Maneva}, {Manganaro}, {Mannheim}, {Maraschi}, {Mariotti},
  {Mart{\'{\i}}nez}, {Masuda}, {Mazin}, {Mielke}, {Minev}, {Miranda},
  {Mirzoyan}, {Moralejo}, {Moreno}, {Moretti}, {Nagayoshi}, {Neustroev},
  {Niedzwiecki}, {Nievas Rosillo}, {Nigro}, {Nilsson}, {Ninci}, {Nishijima},
  {Noda}, {Nogu{\'e}s}, {Paiano}, {Palacio}, {Paneque}, {Paoletti}, {Paredes},
  {Pedaletti}, {Pe{\~n}il}, {Peresano}, {Persic}, {Pfrang}, {Prada Moroni},
  {Prandini}, {Puljak}, {Garcia}, {Reichardt}, {Rhode}, {Rib{\'o}}, {Rico},
  {Righi}, {Rugliancich}, {Saha}, {Saito}, {Satalecka}, {Schweizer}, {Sitarek},
  {{\v S}nidari{\'c}}, {Sobczynska}, {Stamerra}, {Strzys}, {Suri{\'c}},
  {Takahashi}, {Tavecchio}, {Temnikov}, {Terzi{\'c}}, {Teshima},
  {Torres-Alb{\`a}}, {Tsujimoto}, {Vanzo}, {Vazquez Acosta}, {Vovk}, {Ward},
  {Will}, {Zari{\'c}}, {Glawion}, {Takalo}, \& {Jormanainen}}]{tev_magic}
{MAGIC collaboration}, {Ansoldi}, S., {Antonelli}, L.~A., {et~al.} 2018, ArXiv
  e-prints, arXiv:1806.01559

\bibitem[{{Mirzoyan}(2017)}]{mirzoyan2017}
{Mirzoyan}, R. 2017, The Astronomer's Telegram, 9929

\bibitem[{{Mukherjee} \& {VERITAS Collaboration}(2017)}]{mukherjee2017}
{Mukherjee}, R., \& {VERITAS Collaboration}. 2017, The Astronomer's Telegram,
  9931

\bibitem[{{Mushotzky} {et~al.}(1993){Mushotzky}, {Done}, \&
  {Pounds}}]{richard1993}
{Mushotzky}, R.~F., {Done}, C., \& {Pounds}, K.~A. 1993, \araa, 31, 717

\bibitem[{{Nagai} {et~al.}(2010){Nagai}, {Suzuki}, {Asada}, {Kino}, {Kameno},
  {Doi}, {Inoue}, {Kataoka}, {Bach}, {Hirota}, {Matsumoto}, {Honma},
  {Kobayashi}, \& {Fujisawa}}]{nagai2010}
{Nagai}, H., {Suzuki}, K., {Asada}, K., {et~al.} 2010, \pasj, 62, L11

\bibitem[{{Oh} {et~al.}(2018){Oh}, {Koss}, {Markwardt}, {Schawinski},
  {Baumgartner}, {Barthelmy}, {Cenko}, {Gehrels}, {Mushotzky}, {Petulante},
  {Ricci}, {Lien}, \& {Trakhtenbrot}}]{bat_105}
{Oh}, K., {Koss}, M., {Markwardt}, C.~B., {et~al.} 2018, \apjs, 235, 4

\bibitem[{{Reis} \& {Miller}(2013)}]{reis2013}
{Reis}, R.~C., \& {Miller}, J.~M. 2013, \apjl, 769, L7

\bibitem[{{Sanfrutos} {et~al.}(2013){Sanfrutos}, {Miniutti},
  {Ag{\'{\i}}s-Gonz{\'a}lez}, {Fabian}, {Miller}, {Panessa}, \&
  {Zoghbi}}]{sanfrutos2013}
{Sanfrutos}, M., {Miniutti}, G., {Ag{\'{\i}}s-Gonz{\'a}lez}, B., {et~al.} 2013,
  \mnras, 436, 1588

\bibitem[{{Scharw{\"a}chter} {et~al.}(2013){Scharw{\"a}chter}, {McGregor},
  {Dopita}, \& {Beck}}]{scharwachter2013}
{Scharw{\"a}chter}, J., {McGregor}, P.~J., {Dopita}, M.~A., \& {Beck}, T.~L.
  2013, \mnras, 429, 2315

\bibitem[{{Schmidt} {et~al.}(2002){Schmidt}, {Fabian}, \&
  {Sanders}}]{X-ray_isothermal}
{Schmidt}, R.~W., {Fabian}, A.~C., \& {Sanders}, J.~S. 2002, \mnras, 337, 71

\bibitem[{{Sikora} {et~al.}(2007){Sikora}, {Stawarz}, \& {Lasota}}]{sikora2007}
{Sikora}, M., {Stawarz}, {\L}., \& {Lasota}, J.-P. 2007, \apj, 658, 815

\bibitem[{{Suzuki} {et~al.}(2012){Suzuki}, {Nagai}, {Kino}, {Kataoka}, {Asada},
  {Doi}, {Inoue}, {Orienti}, {Giovannini}, {Giroletti}, {L{\"a}hteenm{\"a}ki},
  {Tornikoski}, {Le{\'o}n-Tavares}, {Bach}, {Kameno}, \&
  {Kobayashi}}]{suzuki2012}
{Suzuki}, K., {Nagai}, H., {Kino}, M., {et~al.} 2012, \apj, 746, 140

\bibitem[{{Tanada} {et~al.}(2018){Tanada}, {Kataoka}, {Arimoto}, {Akita},
  {Cheung}, {Digel}, \& {Fukazawa}}]{tanada2018}
{Tanada}, K., {Kataoka}, J., {Arimoto}, M., {et~al.} 2018, \apj, 860, 74

\bibitem[{{Wilkins} \& {Fabian}(2011)}]{wilkins2011}
{Wilkins}, D.~R., \& {Fabian}, A.~C. 2011, \mnras, 414, 1269

\bibitem[{{Wilman} {et~al.}(2005){Wilman}, {Edge}, \& {Johnstone}}]{wilman2005}
{Wilman}, R.~J., {Edge}, A.~C., \& {Johnstone}, R.~M. 2005, \mnras, 359, 755

\end{thebibliography}
\end{document}